\documentclass[11pt]{article}
\def\beq{\begin{eqnarray}}
\def\eeq{\end{eqnarray}}
\def\beqs{\begin{eqnarray*}}
\def\eeqs{\end{eqnarray*}}
\def\dl{\delta}
\def\erf{\mbox{Erf}\>}
\newcommand{\be}{\begin{equation}}
\newcommand{\ee}{\end{equation}}
\newcommand{\lll}{\langle}
\newcommand{\rrr}{\rangle}
\newcommand{\T}{\mbox{Tr}}

\def\centeron#1#2{{\setbox0=\hbox{#1}\setbox1=\hbox{#2}\ifdim
\wd1>\wd0\kern.5\wd1\kern-.5\wd0\fi
\copy0\kern-.5\wd0\kern-.5\wd1\copy1\ifdim\wd0>\wd1
\kern.5\wd0\kern-.5\wd1\fi}}
\def\ltap{\;\centeron{\raise.35ex\hbox{$<$}}{\lower.65ex\hbox{$\sim$}}\;}
\def\gtap{\;\centeron{\raise.35ex\hbox{$>$}}{\lower.65ex\hbox{$\sim$}}\;}
\def\gsim{\mathrel{\gtap}}
\def\lsim{\mathrel{\ltap}}

\begin{document}
\begin{titlepage}
\begin{flushright}
{ITP-UU-01/32}
\end{flushright}

\vskip 1.2cm

\begin{center}

{\LARGE\bf Operator product expansion and confinement.
}

\vskip 1.4cm

{\Large  V.Shevchenko$^{a,b}$, Yu.Simonov$^b$}
\\
\vskip 0.3cm
{\it $^a$Institute for Theoretical Physics, Utrecht University,
Leuvenlaan 4 \\ 3584 CE Utrecht, the Netherlands}
\\
\vskip 0.3cm
{\it $^b$Institute of Theoretical and Experimental Physics,
B.Cheremushkinskaya 25 \\ 117218 Moscow, Russia } \\

\vskip 2cm

\begin{abstract}
Operator product expansion technique is analyzed in abelian and nonabelian 
field theoretical models with confinement. Special attention is paid to the regimes 
where nonzero virtuality of vacuum fields is felt by external currents.
It is stressed that despite the physics of confinement is sometimes considered as being
caused by "soft" fields, it can exhibit the pronounced "hard" effects in OPE.
\end{abstract}
\end{center}

\vskip 1.0 cm

\end{titlepage}

\setcounter{footnote}{0} \setcounter{page}{2}
\setcounter{section}{0} \setcounter{subsection}{0}
\setcounter{subsubsection}{0}


\section{Introduction}

Inclusion of nonperturbative contributions \cite{1} (
proportional to the gauge-invariant local condensates) in
the standard perturbative OPE \cite{2} allowed to formulate a
powerful method of QCD sum rules \cite{1} (for reviews see
\cite{3,4,5}). Nevertheless some questions in the method were
formulated by the original authors \cite{6,7}  and 
still remain unanswered.

In particular, the relation between the 
property of confinement and structure of t
he sum rules series has never been clearly established.  
On the one hand, one could guess that confinement appears as a result of
partial summation of some OPE subseries, while, on the other hand,
confinement itself might introduce some new unconventional 
terms in OPE series, with the structure different from the standard
form. 

The phenomenological  implication of such new terms, e.g.
$O(1/Q^2)$, was investigated in \cite{8}, where is was related
with the short distance nonperturbative physics. 
The authors of \cite{9,10,11,12} checked the role
of confinement for QCD sum rules exploiting nonrelativistic solvable models,
and exact results for Green's
functions were compared to the sum rule results.

Especially popular is the example of nonrelativistic particle in
oscillator potential, with the Euclidean short-time expansion of Green's function (in 2d, for detailed discussion see \cite{5}, cf. the 3d case in
\cite{9})
\be
G^{osc}(T)=\frac{m}{2\pi T} \left(1-\frac{(\omega
T)^2}{6}+\frac{7}{360} (\omega T)^4+...\right) \label{eq1} 
\ee 
Here the first term comes from the free Green's  function while the next
terms play the role of "condensates" namely identifying Borel
mass $\varepsilon=\frac{1}{T}$, one has typical OPE structures:
${\omega^2}/{\varepsilon^2}$ and
${\omega^4}/{\varepsilon^4}$.

The result (\ref{eq1}) has widely been used as an argument that
confinement (i.e. long distance soft physics) cannot modify the
standard OPE and confinement effects should be looked for in the
partial sums of the type  $\sum^\infty_{n=0} c_n(Q^2)\lll D^n
F(0)D^n F(0)\rrr$.

In what follows we shall demonstrate explicitly that confinement
modifies the standard OPE for relativistic quark Green's function:
new terms appear, which bring unusual power terms in OPE.

It will be shown that the expansion (\ref{eq1}) is typical for
nonrelativistic potential Green's functions, while for
relativistic particles in the confining fields (or in the
confining potential) a specific long-distance instability
(divergence) occurs in the perturbative expansion, 
which could lead to new
power terms.

Let us stress from the beginning
an important difference between OPE in coordinate 
and momentum spaces which was discussed already in original papers
\cite{1,1u1} and which will be seen clearly in what follows. 
Studying small $x$-expansion of a product of two operators
$
\lll T\{J(0)J(x)\} \rrr $
when $x\to 0$, one observes that in relativistic case
(contrary to nonrelativistic one) small value of $x$ 
does not confine virtualities of internal 
lines in the coresponding diagram in any way. 
In other words, virtual particles created and annihilated
by operators $J$ can travel over large distances
in coordinate space whichever small $x$ is.  
As a result, in confining theory the product of operators 
taken at two nearby points 
carries information about large-distance behaviour of 
a theory even if $x$ is much smaller than typical
confinement scale $\lambda^{-1}$.

To clarify the mechanism of this phenomenon we start in the next
chapter with the Green's function of relativistic quark in the
linear confining potential of static antiquark, corresponding to
the Dirac equation with scalar linear potential. We shall expand
Green's function in powers of string tension (or equivalently in
powers of Euclidean time $T$) and find explicitly a new dominant
term at small $T$, and estimate other terms.
Comparison with the corresponding nonrelativistic Green's function
is done and demonstrates that no unusual terms appear in the
latter case, the expansion being essentially of the same type as
in (\ref{eq1}). The reason for that is traced to the structure of
the nonrelativistic free Green's function, for which spacial
deflection of particle $\Delta x$ is limited by the time elapsed
$\Delta t$, $\Delta x\sim \sqrt{\frac{\Delta t}{m}}$.

Situation is different however in momentum space.
Large external momentum $Q$ plays a role of infrared
cut-off and if it is much greater than particle mass $m$
and nonperturbative scale
$\lambda$, one can successfully perform systematic expansions
over $m^2/Q^2$ and $\lambda^2 / Q^2$. This is how the standard
OPE technique works. Nevertheless the 
remaining problem in this case is
to determine the structure of the latter, nonperturbative subseries.
The problem here is that in real QCD there are several
different nonperturbative scales. The best known are given by 
 nonperturbative quark and gluon \cite{1} condensates 
$
\lll \bar\psi \psi \rrr$ , $
\left\lll F_{\mu\nu}F_{\mu\nu} \right\rrr$.
One can include in analysis 
higher irreducible condensates as well. 
Another important scale is given by condensate virtualities,
see expressions (\ref{de}), (\ref{dea}) below.
So even remaining in the standard OPE
framework, one can set oneself the task of summation 
of different subseries in full $\lambda^2 / Q^2$--expansion.
It will be seen below how this problem is solved
in particular cases.

Moreover we present a few examples in section 6 
where OPE in momentum space
starts from the terms, which nontrivially
account for (monopole) condensate virtuality and hence 
would be considered as subleading in conventional 
expansion.

The field-theoretical models are discusses in section 3, where the
QCD equations for the heavy-light system
 obtained in the limit of large $N_c$ in \cite{13} are discussed.

 It is shown, in particular, basing on the subsequent results in 
\cite{14},\cite{15},
 that exact equations have a nonlinear kernel, which at large
 spacial distances reduces to the linear confining term $\sigma
 |\vec r |$, and hence the expansion of the Green's function reduces
 to the potential example considered in section 2.

We briefly consider abelian models with confinement in section 
6 such as QED with monopoles and Abelian Higgs model 
and study influence of confinement on short-time behaviour of
Green's functions. 
We also discuss various approaches related to OPE such as 
Feynman-Schwinger proper time method (section 5) and spectral
representations of Green's functions (section 7) and study
interplay between confinement and OPE in these frameworks.
Finally we present short conclusion and outlook.

\section{Relativistic Green's function of a confined quark}

 We study in this chapter Green's  function of the Dirac
 equation in the Euclidean space-time
 \be
 -i(\hat \partial +m+\sigma |\vec{x}|) S(x,y)= \delta^{(4)}(x-y).
 \label{2}
 \ee
 In what follows we shall study the function
 $S(\vec x =0;x_4=0;\vec y =0,y_4=T)\equiv S(T)$ as a function of $T$,
 at small values of $T$.

The free Green's function $S_0(x-y)$ can be written as 
$$
S_0(x)= \int \frac{d^4p}{(2\pi)^4} \frac{\exp(ipx)}{p^2 + m^2} (\gamma p + im) =
i(m-\hat \partial) \lll 0 | (m^2-\partial^2)^{-1} |x \rrr = $$
\be
=i(m-\hat \partial)
\frac{m}{4\pi^2}\frac{K_1(mx)}{x}=i\left( m-\frac{\hat
x}{x}\frac{\partial}{\partial x}\right)
\frac{m}{4\pi^2}\frac{K_1(mx)}{x}\label{3} \ee where $
x=\sqrt{{\vec x}^2+ x^2_4}$ and $K_1$ is the McDonald function. In the
massless limit one obtains
\be
S_0(x)\to \frac{i\hat x}{2\pi^2 x^4}.
 \label{4} \ee
In the first order in $\sigma$ one obtains in the massless limit
\be
S(0,T)=S_0(0,T)+ i\int d^4 x  S_0(0,x)\sigma|\vec x | S_0(x,T)+...
\equiv S_0(0,T) +S_1(0,T)+... \label{5} 
\ee 
where function $S_1$ can be
written in the massless limit as
\be
S_1(0,T)=\frac{i\sigma}{(2\pi^2)^2}\int\frac{\hat x}{x^4} |\vec  x |
\frac{(\hat x-\hat T)}{(x-T)^4} d^4 x \label{6} \ee 
Integration in (\ref{6}) yields
\be
S_1(0,T) =\frac{i\sigma}{8\pi T} \label{7} \ee

Consider now the higher-order terms in the expansion (\ref{5}).
The typical $O(\sigma^n)$ term looks like
\be
S_n(0,T)= i^n \int d^4 x_1 .. d^4 x_n S_0 \sigma|\vec x_1| S_0... \sigma |\vec x_n| S_0 \label{8}\ee 
It is easy to see that the integrals are
infrared divergent at large $|\vec x|$ starting from the term with $n=2$,
however for $m\neq 0$ this divergence is eliminated and integrals
are cut-off by the mass at $x\sim \frac{1}{m}$.
Therefore typical $S_n(0,T)$ has the form for $n>2$ 
\be S_n(0,T)\sim \left (
\frac{\sigma}{m^2}\right )^n m^3  \label{9}\ee 
while the $n=1$ term obtains corrections of the form
\be
S_1(0,T)=\frac{i\sigma}{8\pi T}(mTK_1(mT)+O(mT)).
 \label{10} \ee
It is instructive  to compare  (\ref{4}), (\ref{7}), (\ref{9})
with the nonrelativistic expansion (\ref{eq1}). One can see that
apart from difference in free Green's functions, the first
dynamical term is nonsingular in the nonrelativistic case
(\ref{eq1}), $G_1^{osc}=-\frac{m\omega^2T}{12\pi},$ while it is
singular in relativistic case (\ref{7}) if $T\to 0$.

To clarify the origin of this difference  one can compute
$S_1(0,T)$ for nonrelativistic Green's function with linear
potential. Note, that free Green's function in 3d is
\be
G_0^{nr} (\vec x_1, t_1 ; \vec x_2,t_2)=\left
(\frac{m}{2\pi(t_2-t_1)}\right)^{3/2} \exp \left(
-\frac{m(\vec x_2-\vec x_1)^2}{2(t_2-t_1)}\right) \label{11} \ee a
calculation similar to (\ref{6}) immediately yields
\be
G_1^{nr}(0,T)=\frac{\sigma m}{8\pi} \label{12} \ee which is
nonsingular at small $T$ in contrast to $S_1(0,T)$ in (\ref{7}).
It is easy to see that also all higher terms in $\sigma^n$ are
nonsingular due to the specific feature of nonrelativistic Green's
function (\ref{11}): all time intervals are ordered ($t_n>t_{n-1}
> t_{n-2}$) and all space intervals are cut-off by the time
intervals and the mass, so that quark cannot escape far away during a short
time interval -- in contrast to the relativistic case, when a light
quark can travel as far as $\frac{1}{m}\gg T$ for however small
$T$. Thus crucial difference between nonrelativistic and 
relativistic dynamics causes the different behaviour of the 
Green's functions at small distances/times.

\section{Relativistic equation for the heavy-light system}

In this chapter we shall discuss the situation for the
field-theoretical model, namely for the two-body system
made of a spinor particle with the mass $m$ and heavy
scalar antiparticle whose mass is considered as infinite.
We assume that this "meson" interact with confining gauge-field
background, which is characterized by the Gaussian field strength 
correlator (see review \cite{ddss} and references therein) 
$$
\Delta^{(2)}_{\mu_1\nu_1,\mu_2\nu_2}= \lll {\mbox tr}_c 
(F_{\mu_1\nu_1}(z_1)\Phi(z_1,z_2)
F_{\mu_2\nu_2}(z_2)\Phi(z_2,z_1))\rrr =
  $$
 $$
= \; \frac12\> \left(\frac{\partial}{\partial z_{\mu_1}}
(z_{\mu_2} \delta_{\nu_1 \nu_2} - z_{\nu_2}
\delta_{\nu_1 \mu_2}) +
  \frac{\partial}{\partial z_{\nu_1}}
(z_{\nu_2} \delta_{\mu_1 \mu_2} - z_{\mu_2} \delta_{\mu_1
\nu_2})\right)\> D_1(z_1-z_2) +
  $$
\be  + (\delta_{\mu_1\mu_2} \delta_{\nu_1\nu_2} - \delta_{\mu_1\nu_2}
\delta_{\mu_2\nu_1}) \> D(z_1 - z_2)
\label{eq2}
\ee
where $\Phi(x,y)$ stays for the phase factor
\be
\Phi(x,y) = {\mbox{Pexp}}\left( i\int\limits_x^y A_{\mu}(u) du_{\mu} \right)
\label{ppa}
\ee
The Green's function of such system can be represented as follows
\be
\lll \bar\psi(x) \Phi(x,y) \psi(y)\rrr   =  S_0(x,y) +  S_2(x,y) + ..
\label{ss}
\ee
where $S_0$ is given by (\ref{3}) while the first nontrivial interaction term
has the following form 
\be
{\T} S_2(x,y) = \left\lll \int d^4u\int d^4 w \>\T\> \left( S_0(x,u) i{\hat{A}}(u) S_0(u,w) i{\hat{A}}(w) S_0(w,y)\right) \right\rrr 
\label{s2}
\ee
where $\T = {\mbox tr}_{c}\> {\mbox tr}_{L} $ is a product of traces over color and Lorentz indices. We adopt the Fock-Schwinger gauge condition
with the base point $x_0 = x$: $A^a_{\mu}(u) (u-x_0)_{\mu} =0 $.
In this gauge the Green's function of the heavy particle is proportional to unity while the gauge field propagator takes the form 
\be
\lll {\mbox{tr}}_{c} A_{\mu}(u) A_{\nu}(w) \rrr =
 D(0)\cdot[(u-x)(w-x)\dl_{\mu\nu} - (u-x)_{\nu}(w-x)_{\mu}]\cdot f(u,w)
\label{uu16}
\ee
where dimensionless function $f(u,w)$ is given by the following expression
\be 
f(u,w) = \frac{1}{D(0)}\>\int\limits_0^1 d\alpha \alpha \int\limits_0^1 d\beta \beta
D(\alpha u - \beta w)
\label{prop}
\ee
Functions of this kind are often used in the formalism
of coordinate gauges, one can find in Appendix of the present paper 
detailed analysis of $f(u,w)$ for particular choice of 
Gaussian ansatz $D(z)=D(0)\exp(-z^2/T_g^2)$.  
We shall keep only the function $D(z)$ in what follows since the 
function $D_1(z)$ is not responsible for confinement effects. It was also 
found on the lattice that nonperturbative part of $D_1(z)$ is significantly smaller than
that of $D(z)$ in QCD, see \cite{ddss} and references therein.

In momentum space (\ref{s2}) takes the form:
$$
{\T} S_2(x,y) = 4im\int\frac{d^4 l}{(2\pi)^4} \int\frac{d^4 k}{(2\pi)^4} 
\int\limits_0^1 \alpha d\alpha \int\limits_0^1 \beta d\beta \>\exp(il(x-y))\cdot
$$
$$
\cdot \frac{D(k^2)}{l^2 + m^2}\cdot  {\left[ \frac{\partial}{\partial r_{\rho}} \frac{\partial}{\partial s_{\sigma}}
\right]}_{r=k\alpha \atop s=k\beta} \; \left\{\frac{1}{(l-s+r)^2 + m^2}  \frac{1}{(l-s)^2 + m^2}\cdot \right. 
$$
\be
\left.\cdot (\dl_{\rho\sigma} (3m^2 - l^2 -2lr -sr +s^2) + 4l_{\rho}l_{\sigma}
-4l_{\rho} s_{\sigma} + 2l_{\rho} r_{\sigma} -2l_{\sigma} s_{\rho}
+ 2s_{\rho} s_{\sigma} - r_{\sigma} s_{\rho} - r_{\rho} s_{\sigma} )\vphantom{\frac{1}{(l-s)^2 + m^2}}
\right\}
\label{s3}
\ee

The properties of the expression (\ref{s3}) are determined 
by the interplay of external parameters such as 
particle mass $m$ and distance $|x-y|$ and properties of 
the confining background encoded in the function $D(z)$. 
In case of QCD the latter is usually found on the lattice 
\cite{lat0}. It decays with distance and has some 
typical correlation length scale which we denote as $T_g$ throughout the 
paper. The exact dependence of $D(z)$ on $z$ is of no principal importance, one 
usually takes exponential fits (see \cite{ddss}). At the origin $D(z)$ is 
normalized to the nonperturbative gluon condensate, according to
$$
D(0) = \frac{1}{12}\> \lll {\mbox{tr}}_{c}F_{\mu\nu}F_{\mu\nu} \rrr 
$$
It is worth mentioning that the actual numerical value of $T_g$ in gluodynamics and QCD is rather small: it is estimated as 0.22 Fm for quenched $SU(3)$ and as
0.34 Fm for full QCD with four flavours \cite{lat0,253,ddss}. As it will be clear from what follows
this circumstance bounds region of applicability of conventional OPE
based on local condensates.

We study first the heavy quark case, i.e. we assume that $mT_g \gg 1$.
The integrals in (\ref{s2}), (\ref{s3}) are saturated at momenta $l^2$ of the order of the mass $m^2$ 
and one can make systematic expansion over $1/mT_g$. 
Straightforward although rather lengthy calculation leads to the following answer for the heavy quark condensate:
\be
\T \> S_2 (x,x) = \frac{-i\lll {\mbox{tr}}_{c}F_{\mu\nu}F_{\mu\nu} \rrr 
}{24 \pi^2 m}\> \left[ 1 - \frac{44}{45} \frac{1}{m^2 {\tilde{T}}_g^2} + {\cal O} \left(\frac{1}{ m^4 {\tilde{T}}_g^4}\right) \right]
\label{heavym}
\ee    
where ${\tilde{T}}_g$ is defined as 
\be
\frac{1}{{\tilde{T}}_g^{2}} = \frac{1}{4D(0)}\> \int\frac{d^4k}{(2\pi^4)} D(k^2) k^2 = 
\frac{\lll {\mbox{tr}}_{c}(FD^2F) \rrr }{\lll {\mbox{tr}}_{c} F^2 \rrr }
\label{de}
\ee
where the last relation is valid in Gaussian approximation 
when all contributions from higher correlators are neglected.
Let us mention that virtuality of quark condensate 
usually measured by the quantity
\be
\lambda_q^2 = \frac{\lll \bar\psi D^2 \psi \rrr }{\lll \bar\psi \psi \rrr }
\label{dea}
\ee
in the sum rule approach
is comparable with that of the gluon condensate (\ref{de}),
indeed, $\lambda_q^2 = 0.4 \pm 0.1 {\mbox{GeV}}^2$ according to \cite{fff},
while ${T}_g$ was found on the lattice 
to be $ 0.34\pm 0.02 {\mbox{Fm}}$ in $SU(3)$ with 4 dynamical flavours \cite{253}, 
i.e. $\lambda_q \tilde T_g$ is of the order of one. It could be instructive therefore
to reconcile our approach with the method of nonlocal quark
condensates worked out in \cite{36,37}.

For $D(z)\propto \exp(-z^2/T_g^2)$ with the correlation length $T_g$ (as is used in 
Appendix) one has $T_g = \sqrt{2} {\tilde{T}}_g$.
The first term in the expansion (\ref{heavym}) is well known OPE--result for the heavy quark condensate \cite{con}, 
see also \cite{musa}. 
The second term is the first nonlocal correction. It is worth mentioning that due to the smallness of 
$T_g$ (see above) it can be omitted as numerically small 
correction for $b,t$ quarks only, while for $s,c$ quarks 
keeping only the first term in the expansion in $1/mT_g$ is not to be considered as good approximation.

Equations of the form (\ref{de}), (\ref{dea}) account for
nonzero virtuality of vacuum lines in standard OPE language -- one considers quantum 
averages, which contain derivatives. As we shall see in what follows,
this language is not universal and implicitely assumes small averaged virtuality
corresponding to the vacuum state, i.e. large $T_g$ limit.
Another essential ingredient of this language is the use of equations of motion for such
averages. Although it is rather easy to justify the validity of this component of the approach in 
abelian case, to the best of author's knowledge, this procedure has never been proved
for nonabelian theories with the level of rigour adopted in the field. 
Since we are discussing nonlocal correlators, the following remark is of importance. 
Consider parallel transported field strength tensor $F_{\mu\nu}$,
i.e.
$$
G_{\mu\nu}(x,x_0) = \Phi(x_0,x) F_{\mu\nu}(x) \Phi(x,x_0)
$$  
and nonlocal gauge-invariant two-point correlator 
\be
\lll {\mbox{tr}}_{c} G(x,x_0)G(y,x_0) \rrr
\label{gg}
\ee
The above correlator depends on the positions of the points $x,y,x_0$ and on profiles of 
the contours used in factors $\Phi$. However, if $x\to y$ all these 
dependences disappear (phase factors cancel each other, while $x$-dependence is
prohibited by translational invariance) and the resulting local average coincide with 
$\lll  {\mbox{tr}}_{c} F^2 \rrr $. Let us consider now 
expansion of (\ref{gg}) if $|x-y|$ is small. In principle one might consider
two different expansions, with correlators involving derivatives
in both cases. In the first case it reads:
\be
\lll {\mbox{tr}}_{c} G_{\mu\nu}(x,x_0)G_{\rho\sigma}(y,x_0) \rrr \approx 
\lll  {\mbox{tr}}_{c} F^2 \rrr + (y-x)_{\alpha} \cdot
\left\lll {\mbox{tr}}_{c} G_{\mu\nu}(x,x_0)\left.\left[\frac{\partial G_{\rho\sigma}(y,x_0) 
}{\partial y_{\alpha}} \right]\right|_{y=x}
\right\rrr + ... 
\label{ggq}
\ee
where the derivative is given by 
\be
\frac{\partial G_{\rho\sigma}(y,x_0)}{\partial y_{\alpha}} =
\Phi(x_0, y)\left( D_{\alpha}F_{\rho\sigma}(y) + i(y-x_0)_{\beta}
\int\limits_0^1 sds [G_{\beta\alpha}(z,y), F_{\rho\sigma}(y)] \right) \Phi(y,x_0)
\label{po2}
\ee
and $[..,..]$ in (\ref{po2}) denotes commutator with respect to the color indices.
The second term (and all higher terms) in the r.h.s. of (\ref{ggq}) contains nonlocal part and
depends on contour profiles and on the position of the points $x,x_0$
unless $x=y=x_0$. On the other hand the expansion goes in powers of
the quantity $(y-x)$ which is assumed to be small.

In the second case one expands each $G$ in (\ref{gg}) in the vicinity 
of the point $x_0$:
$$
\lll {\mbox{tr}}_{c} G_{\mu\nu}(x,x_0)G_{\rho\sigma}(y,x_0) \rrr \approx 
$$
$$
\approx
\lll {\mbox{tr}}_{c} (F_{\mu\nu}(x_0)+ (x-x_0)_{\alpha} D_{\alpha}F_{\mu\nu}(x_0)+ ..)
(F_{\rho\sigma}(x_0)+ (y-x_0)_{\alpha} D_{\alpha}F_{\rho\sigma}(x_0)+ ..) \rrr \approx
$$
\be
\approx
\lll  {\mbox{tr}}_{c} F^2 \rrr + (y-x_0)_{\alpha} \cdot
\lll {\mbox{tr}}_{c} F_{\mu\nu}(x_0)D_{\alpha}F_{\rho\sigma}(x_0) \rrr 
+  (x-x_0)_{\alpha} \cdot \lll {\mbox{tr}}_{c} D_{\alpha}F_{\mu\nu}(x_0) F_{\rho\sigma}(x_0)\rrr + ... 
\label{po5}
\ee
This is an expansion adopted in conventional OPE. In contrast with (\ref{ggq})
nonlocal parts are absent, the price to pay 
however is that the expansion goes in $x-x_0$ , $y-x_0$ instead of $y-x$. Needless to say
that in many physical applications $|y-x|$ can be small 
whereas $|x-x_0|$ and $|y-x_0|$ are very large.
Notice also that opposite situation is impossible: smallness
of $|x-x_0|$, $|y-x_0|$ implies smallness of $|y-x|$. 

After this rather academic discussion we come back to the
limit of small quark mass and/or correlation length 
$mT_g \ll 1$, which is opposite to what has been explored 
in (\ref{heavym}). 
As it was already mentioned, in real QCD parameter
$\lll {\mbox{tr}}_{c} F^2 \rrr T_g^4$ can be considered as small, even in presence
of dynamical quarks. In particular, typical momenta $l^2$ in (\ref{s3})
can be rather large in comparison with nonperturbative scale given by the
condensate $\sqrt{\lll {\mbox{tr}}_{c} F^2 \rrr}$ but still small when compared to  
nonlocality scale $\sim T_g^{-2}$. Test particle resolves 
nonlocality of 
vacuum field correlations   
in this regime. 

The Green's function $S_2(x,y)$ we are interested in is defined in (\ref{s2}).
We are working in coordinate representation here and choose the Fock-Schwinger gauge
reference point $x_0$ at the origin $x_0=x=0$. We 
rewrite $S_2(x=0,y)$ using (\ref{uu16})
as
$$
S_2(0,y) = \frac{iD(0)}{64\pi^6}\int d^4 u \int d^4 w \left[
\left( \frac{m}{u^2}  - 2 \frac{\hat{u}}{u^4}  \right)\cdot
(4(uw) - \hat{u}\hat{w}) \cdot \right.
$$
\be
\left.
\cdot f(u,w)\cdot \left(2\frac{\hat{u} - \hat{w}}{(u-w)^4} + \frac{m}{(u-w)^2} \right)
\left(2\frac{\hat{w} - \hat{y}}{(w-y)^4} + \frac{m}{(w-y)^2} 
\right)\right]
\label{s33}
\ee
where we have kept only linear in mass $m$ terms 
in propagators since we consider
small mass limit. The kernel $f(u,w)$ is defined in 
(\ref{prop}) and $\hat u = u_{\mu} \gamma_{\mu}$.

The actual value of this integral is defined by the properties
of the function $f(u,w)$ which encodes all nonperturbative dynamics 
in the chosen Gaussian approximation. They are rather peculiar however
(see Appendix) and this circumstance precludes one to obtain exact analytic answer. 
On the other hand, (\ref{s33}) can be calculated numerically
for any particular ansatz for $D(x)$. Let us investigate general structure
of $S_2$. In massless limit one immediately
obtains $\lim\limits_{y\to 0} S_2(0,y) =0$ according to 
absence of chiral symmetry breaking in the problem in question.
It is seen that $S_2$ is UV-finite (small $u$, $w$ domain) because
nonperturbative background is soft: $\lim\limits_{u,w \to 0} f(u,w) = 1/4$. 
In infrared domain $|u|,|w| \gg T_g$ the integral is convergent 
due to the properties of $f(u,w)$ (see Appendix).
One obtains in massless case the following leading term at small $|y|$:
$$
S_2(0,y) = -i c \cdot D(0)\cdot \hat y + {\cal O}(y^2)
$$
Numerical constant $c$ is determined by  
the function $f(u,w)$, but is $T_g$-independent.
The massive parts of $S_2$ provide finite contribution at $y=0$:
$$
S_2(0,y) \sim imD(0)T_g^2
$$
If mass is increasing and reaches values of the order of $T_g^{-1}$, 
it begins to work as IR cutoff instead of $T_g$ and one comes back
to (\ref{heavym}). However if mass is small then light quarks 
essentially feel the 
virtuality distribution of vacuum gluon fields (i.e. the profile of
$f(u,w)$).

It is instructive to show how the potential problem considered in 
section 2 appears from field-theoretical framework invoking by us here.
To this end one is to consider equation for heavy-light
system which was obtained from the QCD Lagrangian in \cite{13} in
the limit of large $N_c$. Keeping only the Gaussian 
field correlator one has instead of (\ref{2}) the equation for the
quark Green's function (made gauge-invariant due to phase factor
coming from the heavy source propagator)
\be
-i(\hat \partial +m) S(x,y)-i\int M(x,z) d^4 z
S(z,y)=\delta^{(4)}(x-y) \label{13} \ee where the nonlocal kernel
$M(x,z)$ depends on the exact Green's function $S(x,z)$, making
Eq. (\ref{13})  nonlinear. Till the end of this section 
we are working in the so called modified Fock-Schwinger
gauge (see all details in \cite{19}) where the temporal axis
is singled out. We have 
retained for simplicity only color electric part of the correlator
as defined in \cite{20}
 $\lll E_i(x) E_k(z)\rrr\sim \delta_{ik} D(x-z)$. Assuming for 
$D(x-z)$ Gaussian ansatz, one arrives at the
 following form of nonlocal kernel $M(x,y)$,
 \be
 M(x,y)=D(0)(\vec x\vec y ) f(\vec x,\vec y) S(x,y)
 \exp\left(-\frac{(x_4-y_4)^2}{T^2_g}\right)
 \label{15}
 \ee
 where $f(\vec x , \vec y)$ is given in Appendix.
Notice that $\vec x , \vec y$ are three-dimensional vectors here
and not four-dimensional as in (\ref{s33}).
 
 As it was shown in \cite{14} using the relativistic WKB method
 developed in \cite{13}, the function $S(x,y)$ at large $\vec x, \vec y$, i.e.
if $|\vec x|,
 |\vec y|\gg T_g$ can be written in the following form
 \be
 S(h, \vec x, \vec y) =ie^{-(\sigma |\footnotesize{\vec x}| +m) h} g(\vec x, \vec y) \left (
 \begin{array}{ll}
 \theta(h)&\\
 &\theta(-h)
 \end{array}
 \right)
 \label{17}
 \ee
where $h \equiv x_4-y_4$ and $g(x,y)$ is a smeared $\delta$-function
\be
g(\vec x, \vec y)= \tilde{\delta}^{(3)}(\vec x -\vec y),~~
\sigma\vec x^2\sim \sigma \vec y^2\gg 1, \label{18} \ee moreover for
large $\vec x$ and $\vec y$, and $|\vec x-\vec y| \ll |\vec x|\sim |\vec y|$
(see \cite{13} and also Appendix of the present paper)
\be
f(\vec x, \vec y ) \sim \frac{T_g}{|\vec x |} \label{19} \ee
and in this region one can integrate in (\ref{13}) over $d^4 z$,
since
\be
M(x,z)\cong \sigma |\vec x| \delta^{(4)} (x-z) \label{20} \ee 
where the string tension $\sigma = (\pi/2)D(0)T_g^2$ for Gaussian ansatz
(see (\ref{44}) below).
Thus
at large spacial arguments the kernel $M$ coincides with linear
potential considered in the previous chapter. Therefore all
estimates for terms in the expansion proportional to $M^n, n\geq 2
$ are in agreement with those for the local case, eq. (9), since integrals
in these terms are essentially saturated by large spacial
distances, $|\vec x^{(n)}|\gg T_g$.

Although potential behaviour (\ref{20}) is typical for large--$T$ - regime,
it is instructive to show how the nonlocality 
cures $1/T$ behaviour
found in local potential problem. We shall demonstrate now
that Green's functions in question have finite limit
when $T\to 0$ either for small or for large $T_g$.
Let us briefly analyse the situation with the nonlocal equivalent of
(\ref{6}), i.e.
\be
S_1^{(M)}(0,T) =\frac{iD(0)}{(2\pi)^2}\int d^4 x
\int d^4 y \frac{ \hat x(\hat y-\hat T)(\vec x\vec y )f(\vec x, \vec y) S(h, \vec x, \vec y) }{
x^4(y-T)^4}\exp\left(-\frac{h^2}{T_g^2}\right) \label{21} \ee 
with $S$ given in (\ref{17}).
It is convenient to introduce  dimensionless quantities
 \be
 \tilde x_\mu =x_\mu\sqrt{\sigma},~~\tilde y_\mu=
 y_\mu\sqrt{\sigma}, ~~ \tilde T_g=T_g\sqrt{\sigma}, ~~\tilde T=
 T\sqrt{\sigma}.
 \label{22}
 \ee
Rewriting (\ref{21}) in terms of tilde variables, one immediately
realizes, that $S$ tends to 3-dimensional delta-function only in the
limit when $|\tilde x|, |\tilde y|\gg 1$ and otherwise the
integral is defined by the region $|\tilde x|\sim |\tilde y |\sim
1$, when the nonlocality is at work, i.e. $|\tilde x-\tilde y|\sim
|\tilde x|, |\tilde y|.$
Imposing the limit of small $T_g$, i.e.  $\tilde T_g\ll 1$ one
reduces two powers of $\tilde x, \tilde y$ in the numerator of
(\ref{21}), but the integral is still defined by large values of
$\tilde x, \tilde y$ of the order of unity and one finally obtains
\be
S_1^{(M)}(0,T)\sim {\rm const}\cdot \sigma^{3/2} \label{23} \ee 
One
can also show that the same estimate holds true also for higher
terms $O(m^n)$ and write $ S_n^{(M)}(0,T)\sim c_n\sigma^{3/2}.$
Consider now the opposite limit $\tilde T_g\gg 1$, i.e. $T_g\gg
1/\sqrt{\sigma} $. In this case $\tilde T_g$ does not confine the
differences $\tilde x-\tilde y$ in $f(\tilde x, \tilde y)$ and
$\tilde x_4-\tilde y_4$ in the exponent in (\ref{21}) to  small
values as compared to $|\tilde x|, |\tilde x_4|$ or $|\tilde
y|,|\tilde y_4|$.  Therefore the
integration over $d(\tilde x_4-\tilde y_4)$ is limited
only by the exponent in (\ref{17}). As a result one obtais for
$S_1$ the following estimate (as always, we assume mass $m$ to be
not large, $m\lsim \sqrt{\sigma}$) \be S_1^{(M)}(0,T) \sim {\rm const}\cdot
\frac{\sqrt{\sigma}}{T^2_g}
 \label{24} \ee
Thus in both cases the normal procedure of OPE, based on the
analysis of subsequent terms of perturbative expansion (with
separating soft and hard parts of diagrams) is not applicable and
one should sum up the whole series or else solve the nonlinear
equation (\ref{13}) exactly.

\section{Another field-theoretical example: how the linear
confinement is built up out of condensates}

In this chapter we consider another example: a scalar (Higgs-type)
particle interaction with the background Yang-Mills field, and again
calculate the heavy-light Green's function, where light particle
is the color fundamental Higgs, while heavy source is like an
antiquark. The Lagrangian and the Green's function are given by
\be
L=\frac{1}{4}(F^a_{\mu\nu})^2+|D_\mu\varphi|^2-\frac{m^2|\varphi|^2}{2},
~~ G^{(\varphi)}(0,T)=\lll \bar \varphi (0) \Phi(0,T)\varphi
(T)\rrr \label{25} \ee 
One can rewrite  $ G^{(\varphi)}$ as
\be
G^{(\varphi)}(0,T)=\lll
(m^2-D^2_\mu)^{-1}_{0,T}\Phi(0,T)\rrr_B\label{26}\ee
 where we have
introduced the external (background) field $B_\mu$:
$D_\mu=\partial_\mu-ig B_\mu$. For simplicity of consideration we
shall confine ourselves to the vertices
\be
L_4\equiv g^2(B^{ab}_\mu\varphi^b) (B^{ac}_\mu\varphi^c)^+
\label{27} \ee and choose the gauge \cite{19} to write equation
for $G^{(\varphi)}(x,y)$ analogous to (\ref{13}),
\be
(m^2-\partial^2_\mu)G^{(\varphi)}(x,y) + \int I(x,z) G^{(\varphi)}
(z,y) d^4 z=\delta^{(4)}(x-y) \label{28} \ee where
$I(x,z)=I^{(1)}(x,z)+ I^{(2)} (x,z)$, and $I^{(1)}$ refers to the
kernel with one power of  $B_\mu$, while $I^{(2)}$ corresponds to
the Lagrangian (\ref{27}) and can be written as
\be
I^{(2)}(x,y)=\delta^{(4)}(x-y) g^2 B^2_\mu(x) \label{29} \ee
The contributions from $I^{(1)}$ not analyzed by us here are
of the same general structure as that of $I^{(2)}$
apart from nonlocality controlled by the particle mass $m$.
Our consideration in this section is of illustrative purpose
only.

Let us take now the first order graph in $g^2$ (keeping
 only $I^{(2)}$ in (\ref{28})). In Euclidean
 space-time one has
 \be
 G^{(\varphi)}_1(0,T)=g^2\left\lll \int d^4 x\> G^{(\varphi)}_0(0,x) 
B^2_\mu
 (x)G^{(\varphi)}_0(x,T) \right\rrr_{B}
 \label{30}
 \ee
 where
 \be
 G_0^{(\varphi)}(x)=\frac{m}{4\pi^2 x}K_1(m x),~~ x=\sqrt{x^2}.
 \label{31}
 \ee

 We shall be interested in the vacuum averaged expression for
 $G_1^{(\varphi)}$ and to this end one should express $B^2_\mu(x)$
 in terms of  field correlators (one way) or in terms of
 condensates (another -- standard way). In the gauge \cite{19} one can
 write e.g. for $B^2_4$:
 \be
 \lll B^2_4(x)\rrr =\int^x_0 d u_\mu \int^x_0 dv_\nu\lll
 F_{\mu4}(u) \Phi(u,v) F_{\nu4} (v)\rrr
 \label{32}
 \ee
 and using \cite{20} and keeping as in section 3 only the
 confining part $D$, one has
 \be
 \lll B^2_4(x)\rrr =\int^x_0 du_i\int^x_0 dv_i D(u-v),~~ i=1,2,3.
 \label{33}
 \ee
By way of example let us consider exponential ansatz for 
$D(u-v)$. As it was already said, this behaviour of $D(x)$ 
was observed in lattice simulations at distances larger than some typical 
correlation length $T_g$. So one has
\be
 \lll B^2_4(x)\rrr =D(0) \vec x^2 \int^1_0 d\alpha
\int^1_0 d\beta \exp\left(\frac{|\vec x|}{T_g} \cdot |\alpha - 
\beta | \right) 
\label{eoiy}
\ee
Notice absence of additional multipliers 
$\alpha$ , $\beta$ in (\ref{eoiy}) contrary to (\ref{prop}),
it is a property of modified Fock-Schwinger gauge \cite{19}
(temporal axis is singled out) used by us in this section,
instead of usual one, used in (\ref{prop}) (one point is singled
out).

Straightforward integration leads to the following 
result:
\be
 \lll B^2_4(x)\rrr =2D(0) |\vec x | T_g
\left( 1-\frac{T_g}{|\vec x |}\cdot\biggl[ 1-\exp\left(
-\frac{|\vec x |}{T_g}
\right)
\biggr]
\right)
\label{kj}
\ee
At large $|\vec x|\gg T_g$, one has from (\ref{kj}):
  \be
 \lll B^2_4(x)\rrr \approx 2D(0) T_g (|\vec x| - T_g) 
 \label{34}
 \ee
Notice that nonlocality enters (\ref{34}) in explicit way.
At small $|\vec x|$, 
when $|\vec x|/T_g \sim 1$ the linear behaviour of
(\ref{34}) is replaced by the quadratic one 
\be 
\lll B^2_4(x)\rrr
\approx D(0)\cdot \vec x^2 \label{35} 
\ee
It is clear that vacuum field correlator method implemented in
(\ref{32})-(\ref{34}) demonstrates the creation of the string
between the Higgs particle at the point $(\vec x, x_4)$ and static
source at the point $(0,x_4)$.

Now let us look at the same problem from the point of view 
of standard OPE.
According to general
rules \cite{1}, \cite{3}-\cite{5}, one should expand $F_{\mu
4}(u), F_{\nu 4}(v)$ in (\ref{32}) in the vicinity of a point
$\vec u=\vec v=0,u_4=v_4=x_4$ (in the Fock-Schwinger gauge that would
be the point $z_0$ usually chosen at the origin, $z_0=0$). In this
way one obtains \be
 \lll B^2_4(x)\rrr
=\sum_{n,m}\frac{1}{(n+1)!(m+1)!} x_ix_{i_1}... x_{i_n} x_k
x_{k_1}... x_{k_m}\lll D_{i_1}... D_{i_n} F_{i4} (0) D_{k_1}...
D_{k_m} F_{k4} (0)\rrr \label{36} \ee

In this form (\ref{36}) the appearance of the string is not
visible,  and one should rearrange the derivatives in nontrivial way, 
so as to separate out the correlator $D(u-v)$ as in (\ref{33}). Derivatives
of the latter produce powers of $T_g^{-1}$, while dependence on
the sum $\frac12(u_i+v_i)$ in the integral (\ref{33}) is separated
out to yield linear confinement in (\ref{34}).

Now we consider the expansion of $G^{(\varphi)}$ in powers of
$g^2$. From (\ref{34}) and (\ref{30}), (\ref{31}) it is clear that
one obtains
\be
G_1^{(\varphi)} (0,T)\sim g^2\frac{c_4}{m}, ~~ G^{(\varphi)}_n\sim
g^{2n}\left(\frac{c_4}{m^3}\right)^n m^2. \label{37} \ee
where nonlocal constant $c_4 \sim D(0)T^g \sim \int dz D(z)$.
All integrals like (\ref{30}) are diverging at large distances for
$m=0$ and cut-off at $x\sim \frac{1}{m}$ when $ m\neq 0$.

It is instructive to turn to the momentum space
and define the following one-dimensional Fourier-transformed
Green's function:
\be
G_1^{(\phi)}(Q) = \int\limits_{-\infty}^{\infty} d T G_1^{(\phi)}(0,T) 
\exp(-iQT)
\label{ws}
\ee
(Our problem is 3+1 dimensional, we do not perform 4-dimensional 
transformation however
since the temporal axis was separated by our gauge choice
from the beginning).
Since $\lll B_4^2(x) \rrr$ does not depend on the 
temporal coordinate $x_4$, integration in (\ref{30})
is trivial:
\be
G_1^{(\phi)}(M) = \frac{g^2}{16 \pi^2}\int d^3 \vec x
\>\frac{\exp(-2M|\vec x |)}{\vec x^2} \> \lll B_4^2(|\vec x |) \rrr =
\frac{g^2}{16\pi}\> \frac{D(0)}{M^3} \cdot \frac{2MT_g}{1+2MT_g}
\label{yt}
\ee
where $M^2 = Q^2 + m^2$.
Expression (\ref{yt}) has the following asymptotic expansions:
\be
G_1^{(\phi)}(M) = 
\frac{g^2}{16\pi}\> \frac{D(0)}{M^3} \cdot 
\left( 1 - \frac{1}{2MT_g} + \frac{1}{4 M^2 T_g^2} + {\cal O}
\left(\frac{1}{M^3T_g^{3}}\right) \right) \;\; , \;\;\; MT_g \gsim 1
\label{ook1}
\ee
and 
\be
G_1^{(\phi)}(M) = 
\frac{g^2}{8\pi}\> \frac{D(0)T_g}{M^2} \cdot 
(1 - 2MT_g + 4 M^2 T_g^2 +{\cal O} (M^3T_g^3) ) \;\;
, \;\;\; MT_g \lsim 1
\label{ook2}
\ee
It is clearly seen that the actual answer is given
by different series in regions $QT_g \ll 1$ and $QT_g \gg 1$
(we assume that $Q \gg m$ and $M\approx Q$).
The expansion (\ref{ook1}) is associated with the standard
OPE (\ref{36}), while (\ref{ook2}) goes essentially 
in powers of nonlocal quantity. 
It is also worth mentioning
that both expansions (\ref{ook1}) and (\ref{ook2})
are model dependent beyong the leading condensate 
term and actual coefficients in (\ref{ook1}), (\ref{ook2})
are determined by the profile of $D(z)$.

\section{Feynman-Schwinger formalism and OPE}

We start with the same Green's function $G^{(\varphi)}(0,T)$ and
write the Feynman-Schwinger representation (FSR) for it (see
\cite{21},\cite{22} where refs. to earlier  papers are given, for
more discussion see \cite{23}).
\be
G^{(\varphi)} (0,T)=\int^\infty_0 ds \int ({\cal D} z)_{0,T} \exp(-K)\lll
W(C_z)\rrr. \label{39} \ee Here $K=m^2 s+\frac{1}{4} \int^s_0 \dot
z^2_\mu d\tau$, $s$ is Schwinger proper time and $\lll
W(C_z)\rrr$ is a Wilson loop consisting of a straight line
$(0,T)$ and the trajectory of the Higgs particle from $0$ to $T$.
Notice also, that \be ({\cal D}z)_{0T}=\prod^n_{n=1} \frac{d^4\Delta
z(n)}{(4\pi \varepsilon)^2}\int \frac{d^4 p}{(2\pi)^4}
\exp\left(ip\left(\sum\limits_n\Delta z (n)-T\right)\right) \label{40} \ee
In Gaussian approximation 
$\lll W(C_z)\rrr$ 
 can be written as
\be
\lll W(C_z)\rrr =\exp \left(-\frac{g^2}{2} \int_S d\sigma_{\mu\nu} (u)
\int_S d\sigma_{\rho\lambda} (v) \lll F_{\mu\nu} (u)
F_{\rho\lambda} (v)\rrr \right)\label{41} \ee and we have omitted for
simplicity the parallel transporters inside $\lll FF\rrr$. Here
$S$ is the prescribed minimal area surface in the loop $C_z$ 
(there is no sensitivity on the choice of $S$ when all higher
cumulants are kept; with the choice of the Gaussian correlators
and minimal surface, the contribution of all higher correlators
was estimated to be around  few percents, see \cite{ddss} and references 
therein).

For small contour $C_z$ (which means that not only $T$ is small
but also spatial size of the contour is small), one
has from (\ref{41}) \cite{20}
\be
\lll W(C_z)\rrr =\exp\left(-\frac{g^2S^2}{24 N_c}\lll
F^a_{\mu\nu}(0)F^a_{\mu\nu}(0)\rrr \right) \label{42}\ee

For a rectangular contour $C_z$ of an arbitrary size $R\times T$
one can write
\be
\lll W(C)\rrr =\exp\left(-\frac12\int d^2 x \int_S d^2 y D(x-y)\right)
\label{43}\ee
Choosing
for simplicity $D(z)=D(0)\exp (- z^2/T_g^2))$, one has
\be
\sigma =\frac12\int D(z) d^2z= \frac{\pi D(0)}{2}
T^2_g\label{44}\ee and finally
\be
\lll W(C)\rrr =\exp \left(-\frac{\sigma RT}{\pi^2}
L\left(\frac{T^2_g}{R^2}\right) L
\left(\frac{T^2_g}{T^2}\right)\right)  \label{45}\ee where we have
defined
 \be L(u)=\int\limits^\infty_{-\infty} dt\>
e^{-t^2u}\>\frac{\sin^2t}{t^2}\label{46} \ee with the expansions
\be
L(u)=\sqrt{\frac{\pi}{u}}\left \{ 1-\frac{11}{72 u^2}+ \frac{1}{80
u^4} +...\right\},~~ u\gg 1 \label{47}\ee
\be
L(u)=\pi+O(\sqrt{u}),~~ u\ll 1. \label{48} \ee

Now we consider $\lll W\rrr$ inside the integral (\ref{39}). If
one assumes, that for small $T$ one can indeed use the
approximation of small area of the loop $C_z$, i.e. Eq.
(\ref{42}), then one has in the relativistic case, but considering
$T$ small, $T\ll \frac{1}{m}$ and expanding (\ref{42})
\be
G^{(\varphi)} (0,T) = G_0^{(\varphi)} (0,T) + G_1^{(\varphi)}
(0,T)+... \label{49} \ee where
\be
G^{(\varphi)}_1 (0,T)\sim g^2\lll S^2\rrr \lll F^aF^a\rrr
\frac{1}{T^2}\sim g^2 T^2\lll F^aF^a\rrr.\label{50} \ee 
(we have assumed according to what was said before (\ref{42})
that $\lll S^2 \rrr \sim T^4$.
This is
a standard result of OPE analysis with a local condensate
accompanied by higher powers of $T$ (or higher powers of $1/Q^2$
in the momentum representation).

Let us now consider again the term $O(g^2)$, but now taking into
account the nonlocal character of the correlators $\lll F(x)
F(y)\rrr$. To this end we expand the Wilson loop in (\ref{39}) and
making use of two simple identities
$$
({\cal D}z)_{xy} =({\cal D}z)_{xu} d^4 u ({\cal D}z)_{uv} d^4 v({\cal D}z)_{vy} 
$$
\be
\int^\infty_0 ds \int_0^s d\tau_1 \int^{\tau_1}_0 d\tau_2
f(s,\tau_1,\tau_2) =\int^\infty_0 ds \int_0^\infty d\tau_1
\int^{\infty}_0 d\tau_2 f(s+\tau_1+\tau_2,\tau_1+\tau_2, \tau_2)
\label{52} \ee one can write
\be
G_1^{(\varphi)}(0,T)= \int d^4 x \int d^4 y G_0^{(\varphi)}(0,x) \lll A_\mu(x)
A_{\nu}(y)\rrr \dot
x_\mu G_0^{(\varphi)}(x,y) \dot y_\nu
G_0^{(\varphi)}(y,T)\label{53} \ee 
Notation used in (\ref{53})
implies, that $\dot x_\mu(\tau)=\frac{dx_\mu}{d\tau},$ and $\lll
A_\mu(x) A_\nu(y)\rrr$ is expressed through a vacuum average of
field strength $\lll F(u)F(v)\rrr$, e.g. as in the coordinate
gauge \cite{19}
\be
\lll A_4(x) A_4(y)\rrr = \int^x_0 du_i \int^y_0 dv_k\lll F_{i4}
(u) F_{k4} (v)\rrr.
 \label{54} \ee
One can show (see, for example, Appendix B of \cite{23}) that
$ \dot x_\mu \to \stackrel{\leftrightarrow}{\partial}/{\partial
x_\mu}$ and one recoveres in (\ref{53}) the usual
perturbation expansion for $G^{(\varphi)}$, where now in contrast
to the chapter 4 only the linear vertices $\varphi^2A_\mu
\stackrel{\leftrightarrow}{\partial}_\mu$ are taken into account
(the term $\varphi^2A^2_\mu$ would also appear in (\ref{39}) when
one takes into account term with $x=y$).

From (\ref{53}) one can deduce that the r.h.s. stays constant at
large $|\vec x-\vec y|$, while it decreasing for large $|x_4-y_4|$
(this is especially clear when one uses for the correlator
$D(u-v)$ the Gaussian form). Therefore the integral (\ref{53}) is
convergent both at large $x,y$, and at small $x,y$. Integrating
(\ref{53}) one obtains
\be
G_1^{(\varphi)}(0,T) \sim  T_g^2 \lll gF^a gF^a\rrr \sim \sigma,\label{56} \ee 
since $D(0)\sim \lll gF^a gF^a\rrr  \sim
{\sigma}/{T_g^2}$. Thus one obtains a nonlocal constant for
small $T\ll T_g$. Comparing (\ref{50}) and (\ref{56}) one  can see
that at small $T$ the correct (nonlocal) procedure yields
 a larger (dominant) term as compared with the standard OPE
 estimation. The reason again lies in the fact that relativistic
 trajectories occupy larger area for the Wilson loop when treated
 perturbatively and nonlocally, whereas in standard OPE treatment
 one attributes to this term the local condensate, implying  that
 the Higgs particle (or quark) does not go far from the static
 source.

\section{Remarks on OPE in abelian theories with confinement}

We are going to discuss OPE in abelian confining models in this section.
The complications due to path ordering are absent in 
abelian case and one may consider general expression for the 
two--point correlator of the field strengths in the form (\ref{eq2})
where the functions $D(z)$, $D_1(z)$ depend entirely on $z=x-y$.
We assume that confining properties of the theory are caused by 
condensate of monopoles, hence the 
equations of motion take the form:
\be
\partial_{\mu} F_{\mu\nu} = j_{\nu} \;\;\; ; \;\;\; 
\partial_{\mu} {\tilde F}_{\mu\nu}
= J_{\nu} 
\label{eqm}
\ee
where ${\tilde F}_{\mu\nu} = \frac12 \epsilon_{\mu\nu\rho\sigma} F_{\rho\sigma}$ and $j_{\mu}$, $J_{\mu}$ are 
electrically and magnetically charged currents, respectively.
We define polarization operator $\Pi(q^2)$ of the electric currents $j_{\mu}$ as
\be
\int d^d x \lll j_{\mu}(0) j_{\nu}(x) \rrr \exp(iqx) =  
(\dl_{\mu\nu}q^2 - q_{\mu}q_{\nu})\Pi(q^2) 
\label{pppe}
\ee
Differentiating abelian analog of (\ref{eq2}) and taking into account equations of motion, 
it is straightforward to obtain the following relation:
\be
\Pi(q^2) = \int d^d x \left(D(x) + \frac{d}{2}D_1(x) + x^2 \frac{dD_1}{dx^2}\right) \exp(iqx)
\label{ppp}
\ee
In $d=4$ case it can be rewritten in symmetric form as
\be
\lll j_{\mu}(0) j_{\nu}(x) \rrr + \lll 
J_{\mu}(0) J_{\nu}(x) \rrr = 
- \frac16 \>  (\partial^2 \delta_{\mu\nu} - \partial_{\mu}\partial_{\nu}) 
\lll
F_{\alpha\beta}(0) F_{\alpha\beta}(x)\rrr 
\label{ope1}
\ee
where for the condensate one has
\be
\lll
F_{\alpha\beta}(0) F_{\alpha\beta}(x)\rrr =
6(D(x) + {\tilde{D}}(x))
\label{ope3}
\ee
The function ${\tilde{D}}(x) = D(x) + 2D_1(x) + x^2\frac{dD_1}{dx^2}$ 
corresponds to the confining part of the correlator of dual field strengths $\tilde F$ in the same way as $D(x)$ corresponds to the correlator of $F$ 
in (\ref{eq2}).
In case of $d=4$ QED without monopoles, one easily finds \cite{vz}
\be
D(x)\equiv 0 \;\;\; ; \;\;\; D_{1}(x) = \frac{1}{\pi^2} \left(
\frac{e^2(x)}{x^4} - 
\frac{1}{x^2} \frac{d e^2(x^2)}{dx^2} \right)
\label{qed}
\ee
For polarization operator one obtains
\be
\Pi(q^2) = -\frac{1}{\pi^2} \> 
\int d^4 x \exp(iqx) \frac{d^2 e^2(x^2)}{(dx^2)^2}
\label{per}
\ee
It is evident that free field term $\sim e_0^2/x^4$ does not contribute to $\Pi(q^2)$ and the only nonzero contributions to the r.h.s. of (\ref{ppp}) comes from either the running of the charge $e^2(x^2)$ (as in perturbation theory) or from 
nonperturbative parts of $D(x), D_1(x)$, if they are nonzero.  
Let us examine the latter contributions to $\Pi(q^2)$. Standard OPE 
reasoning would suggest to look for leading term of this kind in the form $\lll F^2 \rrr /q^4$. 
It is easy to see that for functions $D(x)$, which are smooth at the 
origin (for example, $D(x)=D(0)\exp(-x^2/T_g^2)$), the corresponding contribution to 
$\Pi(q^2)$ is exponentially suppressed at large $q^2$ (i.e. for $q^2T_g^2 \gg 1 $),
it means that power corrections are absent in this case, in other words nonperturbative 
background is "too soft". In particular, there is no 
$D(0)/q^4$ term.  For $D(x)$ such that it is not smooth but
finite at the origin (e.g. for often used exponential fit $D(x)=D(0)\exp(-|x|/T_g)$, one has as a leading large-$q$ nonperturbative asymptotics
\be
\Delta \Pi(q^2) \sim \frac{D(0)}{T_g}\> {\left(\frac{1}{q^2}\right)}^{\frac52}     
\sim \frac{\lll F^2 \rrr}{T_g q^5}
\ee
The situation becomes even more dramatic if $D(x)$ is singular at the origin 
(as it happens, for example, in the London limit of Abelian Higgs model \cite{dima}), where
$D(x)\sim M^2/x^2 $ if $x\to 0$, one has in this 
particular case
\be
\Delta \Pi(q^2) = \frac{M^2}{q^2 + M^2} \propto \frac{M^2}{q^2}\;\;\; {\mbox{then }} \;\; q^2 \gg M^2      
\label{op9}
\ee
This $1/q^2$ regime in AHM is bounded from above, however, by the Higgs 
mass $m_H$: if $q^2 \gsim m_H^2$, the Ginzburg-Landau description of 
the condensate is not valid, broken symmetry is restored and microscopic degrees of freedom come into play.
Presumably the same reasoning in applicable to "thin" strings scenario, 
proposed in \cite{zz}: at distances much smaller than coherence 
length neither "thick" nor "thin" strings can be formed. 
Notice that the string tension $\sigma$ depends on $m_H$ logarithmically in 
the London limit: $\sigma \sim \log(m_H/M)$.

It is interesting to compare the result (\ref{op9}) with 
an answer for massive photon propagator. It can be obtained 
from (\ref{qed}) taking $e^2(x) = mxK_1(mx)$ which corresponds 
to massive vector field propagator $\lll A_{\mu}(0)A_{\nu}(x) \rrr$.
Differentiation in (\ref{per}) yields
\be
\Pi(q^2) \propto -\frac{m^2}{q^2} \;\;\; {\mbox{then }} \;\; q^2 \gg m^2      
\label{opi}
\ee     
This result is obvious from the form of propagator in momentum
space. Notice the sign difference between (\ref{opi}) and 
(\ref{op9}). It can be said, following \cite{38} that leading 
power correction $\Delta P(q^2)$ in confining theory is caused 
by exchange of massive particle with tachyonic mass.  
This interesting point will be discussed elsewhere.

 \section{Spectral representation of Green's functions and OPE}

 In this chapter we shall look at OPE from another point of view,
 trying to calculate terms of OPE using the known properties of
 spectrum
 of gauge-invariant Green's function.

 This type of analysis was done most extensively for the 'tHooft
 model (1+1 QCD at large $N_c$) \cite{25} where exact results for
 the spectra and Green's  functions are known. (For details of
 analysis the reader is referred to \cite{26,27,28,29,30}. We
 follow most closely notation and   the
line of reasoning of \cite{30}. We consider again the heavy-light
system but now in the $d=1+1$. The Green's function can be written
as
\be
G^{(Qq)}(x)=\lll 0| \bar q (x) \Phi(x,0) q(0)|0\rrr =
\sum^\infty_{n=0} \frac{1}{(2n)!} \lll \bar q (x_\mu
D_\mu)^{2n}q\rrr\label{57} \ee Defining on the other hand the
correlation function
\be
P(q^2) =i\int e^{iqx} d^2x \lll 0| T\{ \bar q Q (x), \bar Qq(0)\}
|0\rrr \sim i \int G^{(Qq)}(x) e^{iqx} d^2 x,\label{58} \ee one
can write and expansion in inverse powers of $E= m_Q- q_0$
\be
P(E)=\frac{1}{E} \left[\lll\bar qq\rrr -\frac{1}{E^2} \lll\bar q
P_0^4q\rrr -...\right] +{\rm pert.part}\label{59}\ee where $P_0=iD_0$.

On the other hand one can write a spectral decomposition
(dispersion relation) for $P(E)$
\be
P(E)=\frac{N_c}{2\pi} m_0\sqrt{\pi} \sum_n\frac{f^2_n}{E+E_n} \sim
\frac{N_c}{\pi} \sum_n \frac{1}{\sqrt{n}
(\sqrt{n}+\varepsilon)}\label{60} \ee 
where we have used notation
$m^2_0\equiv {g^2N_c}/{\pi}, ~~ {2m_0\sqrt{\pi}}\varepsilon=
E$ and relations for  the heavy-light
spectrum \cite{30} \be  E_n=2m_0\sqrt{\pi n} \left
(1+O\left(\frac{\log~n}{n}\right)\right),~ f^2_n=\sqrt{\pi}{n}
\left (1+O\left(\frac{\log~n}{n}\right)\right).\label{61}\ee

Now one can compare (\ref{59}) and (\ref{60}) and expanding the
latter in powers of $\frac{1}{E^n}\sim \frac{1}{\varepsilon^n}$,
one obtains \cite{30} for coefficients in (\ref{59})
\be
\lll \bar q P^{2n}_0q\rrr \sim \lll \bar qq\rrr (\pi m_0)^{2n} n!
\label{62} \ee

The factorial growth of coefficients in (\ref{62}) is typical both
for 1+1 and 3+1, as will be shown below in this chapter.

One can do another derivation of the coefficients (\ref{62})
starting from equations of motion in which case instead of
(\ref{62}) one obtains 
\be
\lll \bar q (x_\mu D_\mu)^{2n} q\rrr\sim x^{2n} n!\lll \bar q
q\rrr\left (-\frac{g^2\lll \bar q q\rrr}{2m_q} \right
)^n.\label{63} \ee

Thus appears another feature (or a puzzle,  as it was formulated
in \cite{30}): condensates computed from the spectrum or from
microscopic equations of motion have drastically different scales:
$m^{2n}_0$ in the first case and $\left
(\frac{m_0^3}{m_q}\right)^{2n}$ in the second case, where $m_q$
tends to zero in the chiral limit.

We shall now show that in the 3+1 QCD at least for $N_c\to \infty$
the situation is very similar to that of the 'tHooft model:

a) OPE coefficients of the $\frac{1}{Q^{2n}}$ expansion
("condensates") have factorially growing behaviour.

b) Condensates calculated from spectrum and from diagrams (plus
equations of motion) are different.

 Consider now the 3+1 problem
-- description of the selfenergy part $\Pi(q^2)$.
For two light quarks the standard OPE of $\Pi(Q^2)$ in the
Euclidean region is well known \cite{1}
\be
\Pi (Q^2) =-\frac{1}{4\pi^2}\left(1+\frac{\alpha_s}{\pi}\right)
\ln\frac{Q^2}{\mu^2} +\frac{6 m^2}{Q^2} + \frac{2m\lll \bar q
q\rrr}{Q^4} + \frac{\alpha_s\lll FF\rrr}{12\pi Q^4} +...
\label{65} \ee

Following \cite{31}
 one can use the background perturbation theory for the
 calculation of $\Pi(Q^2)$ and represent it in the form
 \be
 \Pi (Q^2) =\Pi^{(0)} (Q^2) +\alpha_s\Pi^{(1)}(Q^2)+
 \alpha^2_s\Pi^{(2)}(Q^2).
 \label{66}
 \ee
Let us first consider $\Pi^{(0)}(Q^2)$ (for details of
computations the reader is referred to  \cite{31} and papers
quoted therein).

In the large $N_c$ limit $\Pi^{(0)}(Q^2)$ has the form
\be
\Pi^{(0}(Q^2)=\frac{1}{12\pi^2} \sum^\infty_{n=0}
\frac{C_n}{Q^2+M^2_n}.\label{67}\ee

The masses $M_n$ can be taken as the eigenvalues of the well-known
Hamiltonian, which was derived from QCD with the assumption
of area law for minimal surface and was shown to
be  valid for small angular momentum $L=0,1,2$ \cite{32}, while
for larger $L$ the string rotation should be taken into account,
$\Delta H_{str}$, yielding the correct Regge slope
$(2\pi\sigma)^{-1}$ for masses $M_n$ \cite{32,33,34}
\be
H^{(0)}\Psi_n =M_n\psi_n; ~~H^{(0)}=2\sqrt{{\vec p}^2+m_f^2}+\sigma
r+\Delta H_{str} \label{68} \ee

Solutions to (\ref{68}) can be written in the form
\be
M^2_n= 2\pi\sigma(2n_r+L)+M^2_0\label{69}\ee where $M_0^2\approx
m^2_\rho$. For $C_n$ one has
\be
C_n(L=0)=\frac23 Q^2_fN_cm^2_0,~~C_n(L=2)=\frac13 Q^2_fN_cm^2_0.
\label{70} \ee Here $m^2_0=4\pi\sigma$, and $Q_f$ is the 
electric charge of
quark of flavour $f$.  Taking into account degeneracy of masses
with $L=0, n_r=1$ and $L=2, n_r=0$ the total $C_n$ is the sum \be
\bar C_n=C_n(L=0)+C_n(L=2) =Q^2_f N_c m_0^2.\label{71}\ee

Since $\bar C_n$ does not depend on $n$ in this approximation, one
obtains the sum
\be
\sum^\infty_{n=0}\frac{1}{M^2_n+Q^2}=-\frac{1}{m^2_0}\psi
\left(\frac{Q^2+M^2_0}{m^2_0}\right) +{\rm const} \label{72} \ee
where the constant term is divergent  and is eliminated by renormalization
of $\Pi(Q^2)\to \Pi(Q^2)-\Pi(0)$.

Here $\psi(z)$ is the Euler function
\be
\psi(z)=\frac{\Gamma'(z)}{\Gamma(z)}, ~~ \psi(z)|_{z\to\infty}=\ln
z-\frac{1}{2z}-\sum^\infty_{k=1} \frac{B_{2k}}{2kz^{2k}}\label{73}
\ee where $B_n$ are Bernoulli numbers. Hence at large $Q^2$ the
leading term in (\ref{73}) yields
\be
\Pi^{(0)}(Q^2)=-\frac{Q^2_fN_c}{12\pi^2} \ln
\frac{Q^2+M^2_0}{\mu^2}
+O\left(\frac{m_0^2}{Q^2}\right).\label{74}\ee For $Q^2\gg M^2_0$
this term coincides with the leading term in the OPE (\ref{65})
(the latter is written for $Q_f=1$).

From (\ref{67}) and (\ref{72}) one can compute also the next terms
of the expansion in $\frac{1}{Q^2}$
\be
\Pi^{(0)}(Q^2)=-\frac{Q^2_fN_c}{12\pi^2} \ln
\frac{Q^2+M^2_0}{\mu^2}+\sum^\infty_{n=1}
\frac{\lambda_{2n}m^{2n}_0}{Q^{2n}}
 .\label{75}\ee It is clear
that $\lambda_n$ at large $n$ grow factorially  due to the
asymptotics of Bernoulli numbers,
$B_{2n}=\frac{(-)^{n-1}(2n)!}{2^{2n-1}\pi^{2n}}\zeta(2n).$

Two properties are clearly visible in the expansion (\ref{75})

a) the "condensates" $m^{2n}_0$ are large, $m_0\approx 2.5$ GeV,
as compared to the standard OPE condensates, e.g. $\lll FF\rrr\sim
0.1 \div 0.2$ GeV$^4$

b) the coefficients $\lambda_n$ grow factorially, which is in
agreement with discussion in \cite{28} and analysis of the 'tHooft
model in \cite{29,30}, signifying that the OPE series is
asymptotic.

Thus in both cases, 1+1 and 3+1, when confinement is present and
spectrum contains nondecreasing probabilities $C_n$ (which is the
feature of linear confining interaction) the OPE is a factorially
diverging series, implying renormalon singularities in the Borel
plane \cite{31}. Another feature which is general to both 1+1 and
3+1 theories, is the mismatch between condensates calculated via
spectrum (as in (\ref{75}) and via diagrammatic analysis (as in
(\ref{65})). In \cite{30} a possible solution of this masmatch for
the 1+1 case was suggested, which introduces the notion of
"effective condensates", which may differ from actual condensates
(defined, for example, on the lattice) due to the asymptotic character of
the OPE series.

In 3+1 case there is another possibility to explain the
mismatch, namely one should take into  account that coefficients
$\lambda_n$ of all higher condensates get contribution not only of
the leading terms in $n$ of $M_n$ and $c_n$, but also subleading
terms, and the final result for say $\lambda_4$ could be two
orders of magnitude smaller due to  cancellation  between
different terms, thus removing the mismatch. However this requires
a mechanism of fine tuning between the subleading coefficients,
the  physical reason for which is still not known.

One could leave discussion of the  mismatch at this point, if
another check were not possible. Indeed, let us take the  OPE with
large (spectral) condensates and  do a sum  rule analysis of
experimental data for $e^+e^-\to$ hadrons with $I=1$ 
(see \cite{35}).

This analysis was done in \cite{31} using the hadronic ratio
$R^I(s)=12\pi  {\mbox{Im}} \Pi^I(S)$. For $I=1$ adding the perturbative
terms with known coefficients as in \cite{1, 35}, but taking the
background modified  coupling constant \cite{31} e.g. in one loop
$$\alpha_B(Q^2)=\frac{4\pi}{b_0\ln
\left(\frac{Q^2+M^2_B}{\Lambda^2_B}\right )}
$$
where $M_B\approx 1.5 $GeV, $\Lambda^{(3)}_B\approx 482$MeV,
one has 
\be
R^{I=1}(s) =\frac32 \sum^\infty_{n=0} C_n^{I=1} \delta
(s-M^2_n)+\frac32\left(1+\frac{\alpha_B(s)}{\pi} +1.64
\left(\frac{\alpha_B}{\pi}\right)^2\right)\label{76}\ee
$$C^{I=1}_n=m^2_0,~~ M^2_n= m^2_\rho+n m^2_0,~~ n=1,2,...;
~~C_0=\frac23m^2_0,$$ and the corresponding Borel transform is
\be
\tilde I_0(M)=\frac{2}{3 M^2}\int^\infty_0 ds e^{-s/M^2}
R^{I=1}(s). \label{77}\ee
Substituting (\ref{76}) into (\ref{77}) yields $$ \tilde
I_0(M)=\frac{m^2_0}{M^2}\left\{\frac23 e^{-m^2_\rho/M^2}
+\sum^\infty_{n=1} e^{-(m^2_\rho+nm^2_0)/M^2}\right\}$$ \be
+\frac{\alpha_B(M)}{\pi}+2.94
\left(\frac{\alpha_B(M)}{\pi}\right)^2,~~ m^2_0=4\pi\sigma.
\label{78}\ee

This should be compared to the standard result \cite{1} with
standard (small) condensates $$ \tilde
I^{st}_0(M)=1+\frac{\alpha_s(M)}{\pi}+2.94
\left(\frac{\alpha_s(M)}{\pi}\right)+$$ \be
+\frac{\pi^2}{3}\frac{G_2}{M^4}+\frac{448 \pi^3\alpha_s}{81}
\frac{|\lll 0|\bar q q|0\rrr|^2}{M^6}.\label{79}\ee In (\ref{79})
$\alpha_s(M)$ is standard, i.e. obtained from $\alpha_B$ by
setting $ M_B\equiv 0$.

It is clear  that  (\ref{78}) contains in the Borel plane a set of
poles at $M^2= M^2_k= \pm i \frac{m^2_0}{2 \pi k}, ~~ k= 1,2,...$
and an essential singularity at $M=0$. These features imply
presence of renormalons and are connected to the factorial growth
of coefficients $\lambda_{2n}$ in (\ref{75}).

Now remarkably both Borel transforms lie inside the corridor of
experimental errors, thus describing satisfactorily data with very
different values of condensates (for details of comparison see
\cite{31}). Thus situation is becoming even more unclear: not only
one has two sets of condensates (and consequently two sets of  sum
rules) but in addition experimental data cannot give preference to
one of them.

While leading perturbative large-$M$ asymptotics of $\tilde I_0(M)$ and $\tilde
I_0^{st}(M)$ coincide, there is an important difference 
at small $M$: while $M^2\tilde I_0(M)$
is defined for all $M$,  $M^2\tilde I_0^{st}(M)$ is diverging for
$M\to 0$ due to higher condensates and higher powers of
$\alpha_s(M)$.

\section{Conclusions and outlook}

The main emphasis of the present paper is   the influence of
confinement on the  behaviour of Green's functions in their
dependence on momentum and the  behaviour of  Borel transforms. We
stressed above everywhere the importance of large distances
working in coordinate representation,
especially for light quarks in presence of confinement. As a first
and most clear example the Green's function of Dirac equation with
linear scalar potential was considered and it was demonstrated
that the Eucliden time expansion (equivalent of Borel transform
for heavy-light systems) looks completely different from the
nonrelativistic case, and from the template oscillator Green's
function. In this way it was shown that large distances may be
important even for small Euclidean times and bring about new terms
in the OPE in coordinate space.

As a second example we have treated the nonlinear equations for a
quark in the heavy-light system --nonlocal equivalent of the Dirac
case, and found that again the result is different from what one
would expect in standard OPE, but the terms of expansion turn out
to  be constant, $S_1(T)\sim const\cdot \sigma^{3/2}$. Translating
this contribution into the form of the usual correlation function
$\Pi(Q^2)$ of vector currents (like it is done in the reaction
$e^+e^-\to$ hadrons) one would have the contribution $\Delta
\Pi(Q^2)\sim \frac{m\sigma^{3/2}}{Q^4},$ which is similar to the
standard term $\frac{m\lll \bar q q\rrr}{Q^4}$, and is presumably
one term in the subseries generating $m\lll \bar q q\rrr$. In this
example  large distances, explicitly accounting for in our
analysis, do not produce new OPE terms but give some path to
calculating chiral condensate through confinement  characteristics
(i.e. string tension $\sigma$). 

In section 4, in contrast to that, another problem was elucidated:
how linear confinement is built up out of higher condensates of
OPE, and the answer is given by comparison of Eqs.
(\ref{33})-(\ref{34}) and (\ref{36}). Indeed, the infinite sum of
derivatives of field correlators in (\ref{36}) is equivalent to
the linear confinement term in (\ref{34}), and to extract it
explicitly one needs to rearrange all derivatives.

We have analyzed abelian confining models in section 6
and described different possible sources of nonstandard OPE 
terms, e.g. $1/q^2$.

Finally, the last problem considered in the paper concerns the
derivation of OPE from the spectral representation of the meson
Green function. When the spectrum and coefficients $c_n$
(equivalent of quark decay constants $f_\pi$) are known, the  OPE
is calculated automatically and can be compared with that obtained
"microscopically"- i.e. via Feynman diagrams in the external
fields and equations of motion.

In the $d=1+1$ QCD this program was fully investigated in a series
of papers (see e.g. \cite{30} and refs. therein) and a mismatch
between condensates obtained in those two ways was found.

In the 3+1 QCD situation is similar and as shown in \cite{31} and
in the present paper, the mismatch of condensates in scales and
order of magnitude also is evident. The situation is sharpened by
the fact, that the QCD sum rules for $e^+e^-\to$ hadrons reproduce
experimental data for both choices of condensates.

We have not tried here to resolve this puzzle, and leave it for
the future. There are two important topics in OPE we have not
discussed. First, this is the partial summation of the OPE terms
which can be done by introduction of nonlocal condensates in OPE,
initiated and studied in \cite{36,37}. It would be interesting to
find the link between our treatment of long-distance
nonperturbative physics and the method of nonlocal condensates
worked  out in \cite{36,37}.

Second, the problem of perturbative-nonperturbative interference,
which may produce new singular OPE terms, like $1/Q^2$, which was
discussed in \cite{38,39,8}, 
is touched in section 6 only briefly. This set of problems 
certainly deserves futher study.

\section{Acknowledgments}
The authors acknowledge the support from the grants 
RFFI-00-02-17836, RFFI-00-15-96786 and from the grant 
INTAS 00-110. V.Sh.  
acknowledges the support of the foundation
"Fundamenteel Onderzoek der Materie" (FOM), which is financially
 supported by the Dutch National Science Foundation (NWO).
V.Sh. also acknowledges the support from the grant RFFI-01-02-06284.
Yu.S. acknowledges financial support from INTAS grant 00-00366.

\newpage

\appendix

\section{Appendix}
\setcounter{equation}{0} 
\def\theequation{A.\arabic{equation}}

We discuss in this Appendix the properties of the kernel (\ref{prop}),
which we used in the main text.
The reader is referred to the Appendix 3 of the paper \cite{13} where 
$3d$ counterpart of 
(\ref{prop}) was analysed. Since the kernels of the form (\ref{prop}) play important 
role in the discussed formalism we present an independent 
detailed analysis here both for possible future 
applications and for reader's convenience. 

We are interested in the properties of the following function:
\be
f(\vec \eta, \vec \rho) = \int\limits_0^1 d\alpha \alpha \int\limits_0^1 d\beta \beta \exp\left(-\frac{(\alpha\vec \eta - \beta\vec \rho)^2}{T_g^2}\right)
\label{ap1}
\ee 
where $\vec \eta , \vec \rho$ are $d$-dimensional vectors with angle $\theta$ between them. We denote absolute values of the arguments as $\eta = |\vec \eta |$, $\rho = |\vec \rho |$ and assume in what follows without loss of generality that $ \eta \ge \rho$. The symmetry of formulas below 
with respect to the exchange $\rho \leftrightarrow 
\eta$ (which is manifest in the definition (\ref{ap1})) is 
to be restored by replacements 
$\rho \to {\mbox{min}\> \{ \rho, \eta \} }$ and
 $\eta \to {\mbox{max}\> \{ \rho, \eta \} }$.    
  
It is instructive to consider four different asymptotic regions: 

{\bf 1).} ${\eta} , {\rho} \sim T_g \;\;\;$ {\bf 2).} ${\eta} \gg T_g ; 
{\rho} \sim T_g \;\;\;$ {\bf 3).} ${\eta} , {\rho} \gg T_g ; \theta \gsim 
1
\;\;\;$
{\bf 4).} $\eta , \rho \gg T_g ; \theta \ll 1$

In the region {\bf 1)} one can expand (\ref{ap1}) in Taylor series with respect to both arguments, subsequent integration is straightforward:
$$
f(\vec \eta, \vec \rho) =  \frac14 - \frac{\rho^2 + \eta^2}{8T_g^2} + \frac{2\rho\eta\cos\theta}{9T_g^2} +
 \frac{\rho^4 + \eta^4}{24T_g^4} +   
\frac{\rho^2\eta^2}{8T_g^2}\left(\cos^2\theta + \frac12 \right)
$$
\be
- \frac{2\rho\eta\cos\theta}{15T_g^4}\left(\rho^2 + \eta^2\right) +  
{\cal O}(T_g^{-6})
\label{ap2}
\ee
In derivation of expression (\ref{heavym}) in the main text we have used in fact the leading term of this asymptotics (i.e. 1/4).

In the regions {\bf 2)}, {\bf 3)}, {\bf 4)} we will systematically omit exponentially small terms, i.e. terms proportional to 
$\exp(-\eta^2/T_g^2)$ and also terms $\sim \exp(-\rho^2/T_g^2)$ in the regions {\bf 3)} and {\bf 4)}.
One easily obtains the following expression in the region {\bf 2)} :
\be
f(\vec \eta, \vec \rho) =  \frac{T_g^2}{\eta^2}\left( \frac14 + \frac{\sqrt{\pi}}{6} \frac{\rho\cos\theta}{T_g} + \frac18 \frac{\rho^2}{T_g^2} (2\cos^2\theta -1) - \frac{\sqrt{\pi}}{10} \frac{\rho^3\cos\theta\sin^2\theta}{T_g^3} + {\cal O} (\rho^4) \right)
\label{ap3}
\ee

Now we come to the regions {\bf 3)} and {\bf 4)}. It is instructive to introduce the following variables: 
\be
s =  \frac{(\vec \eta \rho - \vec \rho \eta )^2}{T_g^2} = \frac{4\eta^2  \rho^2}{T_g^2} \> \sin^2\frac{\theta}{2}  \;\;\; ; \;\;\;
q=  \frac{(\vec \eta \rho + \vec \rho \eta )^2}{4 T_g^2} = \frac{\eta^2  \rho^2}{T_g^2} \> \cos^2\frac{\theta}{2} 
\label{ap4}
\ee
and also $\xi = \sqrt{s}/\sqrt{q} = 2 \tan \frac{\theta}{2}$.
  
In the region {\bf 3)} the upper limit of the integration in (\ref{ap1}) can be shifted to infinity up to exponentially small corrections. 
The function $f(\vec \eta, \vec \rho)$ in the region {\bf 3)} can be 
written therefore as
\be
f(\vec \eta, \vec \rho) =  \frac{T_g^4}{4\eta^2\rho^2}\cdot \phi(\xi)
\label{ap21}
\ee
where $\phi(\xi)$ is given by
\be
\phi(\xi) = \frac{\sqrt{\pi}}{8} \frac{(4+\xi^2)^2}{\xi^3} \; 
\int\limits_0^{\infty} dy \exp(-y^2) \left[ \left(1- \erf\left(\frac{y\xi}{2}\right)\right)\left(1-\frac{y^2\xi^2}{2}\right) +
\frac{y\xi}{\sqrt{\pi}}\>\exp\left(-\frac{y^2\xi^2}{4}\right)\right]
\label{ap22}
\ee
The function $\phi(\xi)$ is a monotonically decreasing function of $\xi$.
When $\xi$ is going to infinity, $\phi(\xi)$ is approaching the following asymptotics:
\be
\phi(\xi) = \frac13 + \frac{32}{15}\frac{1}{\xi^2} + {\cal O}
(\xi^{-4}) 
\label{ap23}
\ee
At the point $\xi=2$, which corresponds to $\theta = \pi /2$ and hence orthogonal vectors $\vec \eta$ and $\vec \rho$ one finds $\phi(2)=1$, in agreement with simple direct calculation from (\ref{ap1}).

We are now in the position to analyse the properties of $f(\vec \eta, \vec \rho)$ in the region {\bf 4)}, where $\xi \sim \theta \ll 1$. Making the change of variables, one gets from (\ref{ap1}):
$$
f(\vec \eta, \vec \rho) = -\frac{T_g^2}{\sin\theta} \> \left[\left\{ \int\limits_{ \frac{\sqrt{s}}{\eta \xi} }^0 dy \int\limits_{\frac{y\xi}{2} }^{\frac{\sqrt{s}}{\eta} - \frac{y\xi}{2}} dx +
 \int\limits_0^{-\frac{\sqrt{s}}{\rho \xi}} dy \int\limits_{-\frac{y\xi}{2}}^{\frac{\sqrt{s}}{\rho} + \frac{y\xi}{2}} dx +
\int\limits_0^{-w} dy \int\limits_{-\frac{y\xi}{2}}^{\frac{\sqrt{s}}{\eta} - \frac{y\xi}{2}} dx - 
\right.\right.
$$
\be
\left.\left.
\int\limits_0^{-w}dy \int\limits_{-\frac{y\xi}{2}}^{\frac{\sqrt{s}}{\rho} + \frac{y\xi}{2}} dx \right\} \left( \frac{x^2}{s} - \frac{y^2{\xi}^2}{4s}\right)
\exp\left(-x^2 - y^2\right) \right]
\label{ap6}
\ee 
where $w=\sqrt{q}(1/\rho - 1/\eta) = (\eta - \rho)\cos(\theta/2) / T_g$.
One can rewrite (\ref{ap6}) in the following form
\be
f(\vec \eta, \vec \rho) = \frac{T_g^2}{\sin\theta }\>\frac{1}{s} \> \left[ g\left(\frac{\sqrt{s}}{\eta}, \xi\right) + g\left(\frac{\sqrt{s}}{\rho} , \xi\right) +
f_2 \left(\frac{\sqrt{s}}{\eta}, w,  \xi\right) - f_2\left(\frac{\sqrt{s}}{\rho}, w, -\xi\right)\right]
\label{ap8}
\ee
where the $\xi$-expansion of the functions $g, f_3, f_4$ can be performed systematically. It gives
\be
g(z, \xi) = \frac{\pi}{8} \kappa(z) - \frac{\exp(-z^2) z^2}{4} \cdot \xi -
\frac{\pi}{32} \left( \kappa(z) + \frac{2z^3}{\sqrt{\pi}}\exp(-z^2)\right)\cdot \xi^2 +{\cal O}(\xi^3)
\label{ap10}
\ee
\be
f_2(z,w,\xi) = \erf(w)\frac{\pi}{8} \kappa(z)
 + [1 -\exp(-w^2)] \frac{\exp(-z^2) z^2}{4} \cdot \xi  + {\cal O}(\xi^2)
\label{ap11}
\ee
where the function $\kappa(z)$ is defined as
\be
\kappa(z) =  \erf(z) - \frac{2z}{\sqrt{\pi}}\exp(-z^2)
\label{ap33}
\ee
Extracting coefficient functions in front of higher powers in $\xi$ is a matter of straightforward algebra.

The  expressions (\ref{ap10}), (\ref{ap11}) are exact at the given order in $\xi$ up to omitted exponentially small terms. 
They can be simplified in different limiting cases. If $w=0$ (i.e. $\eta = \rho$ ), one has 
$f_2 = 0$, while the first two terms in the r.h.s. of 
(\ref{ap8}) are equal. In the opposite limit $w\to\infty$ 
the following relations hold true:
\be
\lim\limits_{w\to\infty} \left[
 g\left(\frac{\sqrt{s}}{\rho}, \xi\right) -
 f_2\left(\frac{\sqrt{s}}{\rho}, w, -\xi\right) \right] =0 
\;\;\; ; \;\;\;
\lim\limits_{w\to\infty}
f_2 \left(\frac{\sqrt{s}}{\eta}, w, \xi \right) =
g\left(\frac{\sqrt{s}}{\eta}, -\xi \right)
\ee

Notice that in all cases the first argument of the functions $g, f_2$ need not be small: $\sqrt{s}/\eta = (2\rho/T_g)\sin(\theta/2)$ and in the region {\bf 4)} $\rho \gg T_g$, but $\theta \ll 1$.

In terms of the original variables $\eta, \rho, \theta$ the leading 
term in (\ref{ap8}) can be represented as
\be
f(\vec \eta, \vec \rho) \approx \frac{\pi}{64\sin^3
\frac{\theta}{2}\cos\frac{\theta}{2}}\>\frac{T_g^4}{\eta^2\rho^2}\>
\left[ \kappa\left(\frac{2\rho\sin\frac{\theta}{2}}{T_g}\right) \left(1+\erf(w)\right) + \kappa\left(\frac{2\eta\sin\frac{\theta}{2}}{T_g}\right) \left(1-\erf(w)\right)
\right]
\label{ap14}
\ee
where $w = (\eta - \rho)\cos(\theta/2) / T_g$ and $\kappa(z)$ is defined in (\ref{ap33}), $\kappa(z) > 0$ if $z>0$. This expression is valid in small $\theta$-limit.

Notice that $f$ is non-singular if $\theta\to 0 $ limit (which is evident from
(\ref{ap1})):
\be
\lim\limits_{\theta\to 0} f(\vec \eta, \vec \rho) = \frac{T_g \sqrt{\pi}}{6}\left[ \frac{\rho}{\eta^2} + \frac{\eta}{\rho^2} + \erf\left(\frac{\eta -\rho}{T_g}\right) \left(\frac{\rho}{\eta^2} - \frac{\eta}{\rho^2}\right)\right]
\label{ap34}
\ee

One needs some simple extrapolating representation of (\ref{ap1}) for  practical calculations. Notice that it is only asymptotic behaviour of $f(\vec \eta , \vec \rho)$ that matters, the particular form of Gaussian kernel was taken in (\ref{ap1}) just as an example.
A possible expression respecting all desired properties of $f$ in the regions of large $\eta , \rho$ is as follows:
\be
f(\vec \eta, \vec \rho) \approx \frac{T_g^4}{4 \eta^2 \rho^2} \> l(\theta)
\label{ap56}
\ee 
where the function $l(\theta)$ has the following "focusing" property:
being integrated with a regular function $F(\theta)$, it acts like a 
smoothed $\dl$-function (see \cite{13}):
\be
\int d\theta F(\theta) l(\theta) \approx  
c_1 \frac{\rho^3}{T_g^3}\> F\left( \frac{c_0 T_g}{\rho}\right) + c_2 \int d\theta F(\theta)
\label{ap57}
\ee
where $c_0 , c_1 , c_2 $ are some constants of the order of unity. It is worth reminding that $\rho$ is
the length of the smaller vector in our notation, i.e. $\rho = {\mbox{min}\> \{ \rho, \eta \} }$. In particular, it is seen that in the limit of large 
$\rho \gg T_g$ the small $\theta$ asymptotics gives dominant contribution
unless  $F(\theta)$ vanishes at the origin faster than $\theta^3$.

\end{document}